\documentclass[prd,amsfonts,amssymb,noshowpacs,nofootinbib,preprintnumbers]{revtex4}
\usepackage{amsmath,graphicx,hyperref,epsfig}
\usepackage{color,ulem}

\begin{document}

\title{Cosmological perturbations in mimetic Horndeski gravity}
\author{
Frederico Arroja$^{1}$\footnote{arroja@phys.ntu.edu.tw}, Nicola Bartolo$^{2,3,4}$\footnote{nicola.bartolo@pd.infn.it}, Purnendu Karmakar$^{2,3,5}$\footnote{purnendu.karmakar@pd.infn.it} and Sabino Matarrese$^{2,3,4,6}$\footnote{sabino.matarrese@pd.infn.it}
}
\affiliation{
{}$^{1}$ Leung Center for Cosmology and Particle Astrophysics, National Taiwan University, No.1, Sec.4, Roosevelt Road, Taipei, 10617 Taipei, Taiwan (R.O.C)
\\
{}$^{2}$ Dipartimento di Fisica e Astronomia ``G. Galilei'', Universit\`{a} degli Studi di
Padova, via Marzolo 8,  I-35131 Padova, Italy
\\
{}$^{3}$ INFN, Sezione di Padova, via Marzolo 8,  I-35131 Padova, Italy
\\
{}$^4$INAF - Osservatorio Astronomico di Padova, Vicolo dell'Osservatorio 5, I-35122 Padova, Italy
\\
{}$^5$Institute of Theoretical Astrophysics, University of Oslo, N-0315 Oslo, Norway
\\
{}$^{6}$ Gran Sasso Science Institute, INFN, Viale F. Crispi 7,  I-67100 L'Aquila, Italy
}

\begin{abstract}
We study linear scalar perturbations around a flat FLRW background in mimetic Horndeski gravity. In the absence of matter, we show that the Newtonian potential satisfies a second-order differential equation with no spatial derivatives. This implies that the sound speed for scalar perturbations is exactly zero on this background. We also show that in mimetic $G^3$ theories the sound speed is equally zero. We obtain the equation of motion for the comoving curvature perturbation (first order differential equation) and solve it to find that the comoving curvature perturbation is constant on all scales in mimetic Horndeski gravity. We find solutions for the Newtonian potential evolution equation in two simple models. Finally we show that the sound speed is zero on all backgrounds and therefore the system does not have any wave-like scalar degrees of freedom.
\end{abstract}


\date{\today}
\maketitle

\section{Introduction\label{sec:INT}}

Recently, mimetic gravity (called mimetic dark matter) \cite{Chamseddine:2013kea}, a modification of General Relativity (GR) leading to a scalar-tensor type theory, has attracted considerable attention in the cosmology community. The main reason is that the theory possesses some very attractive features. For example, it was shown that the original theory (called mimetic dark matter) \cite{Chamseddine:2013kea} contains an extra scalar mode (of gravitational origin) which can mimic the behaviour of cold dark matter even in the absence of any form of matter. Soon after it was realised that with a small generalisation of the original theory the scalar mode could be used to mimic the behaviour of almost any type of matter and in this way one can have almost any desired expansion history of the universe \cite{Chamseddine:2014vna}.

The mimetic scalar field was introduced in GR by doing a non-invertible conformal transformation in the Einstein-Hilbert action of the type $g_{\mu\nu}=-w\ell_{\mu\nu}$, where the physical metric is $g_{\mu\nu}$, the auxiliary metric is $\ell_{\mu\nu}$, $w$ is defined in terms of a scalar field $\varphi$ as $w=\ell^{\mu\nu}\partial_\mu\varphi\partial_\nu\varphi$ \cite{Chamseddine:2013kea,Chamseddine:2014vna,Barvinsky:2013mea}. Soon after it was realized \cite{Deruelle:2014zza} that the type of metric transformation that leads to mimetic gravity can be further generalised from the previous transformation to include also a disformal term \cite{Bekenstein:1992pj} as $g_{\mu\nu}=A(\varphi,w)\ell_{\mu\nu}+B(\varphi,w)\partial_\mu\varphi\partial_\nu\varphi$, where $A$ and $B$ are free functions of two variables and they must obey some conditions (see \cite{Bettoni:2013diz} and also \cite{Zumalacarregui:2013pma} where the conditions for disformally coupled theories to have a so-called Jordan frame were discussed) so that the Lorentzian signature is preserved, the transformation is causal and regular, $g^{\mu\nu}$ exists and $A$ and $B$ are related as $B=-A/w+b$, where $b$ is an arbitrary function of $\varphi$ only and it should not cross zero. If $A$ and $B$ are arbitrary functions and do not obey the previous relation then the equations of motion that one obtains are just Einstein's equations \cite{Deruelle:2014zza}.

The stability of mimetic gravity against negative energy states, i.e. ghosts, was studied in \cite{Barvinsky:2013mea}, where it was shown that ghosts are absent if the energy density of the effective fluid is positive.
Ref. \cite{Golovnev:2013jxa} (see also \cite{Barvinsky:2013mea} and \cite{Chamseddine:2014vna}) showed that the original mimetic gravity can be derived from an action with a constraint imposed by a Lagrange multiplier without the need to invoke a disformal transformation. In this paper we will follow this complementary approach of using a Lagrange multiplier.

Recently in \cite{Arroja:2015wpa}, we have generalised some of the previously mentioned results starting from a very general scalar-tensor theory of gravity
(which includes, e.g., Horndeski models~\footnote{
See \cite{Horndeski:1974wa} for the original reference on the Horndeski theory or for example \cite{Deffayet:2011gz,Kobayashi:2011nu} and references therein for a modern formulation.
} and the so-called $G^3$ theories~\cite{Gleyzes:2014dya,Gleyzes:2014qga}). We have shown that such general theories are invariant under the generalised disformal transformations introduced above. However, for a small subset of those transformations, when they are not invertible, the resulting theory is a generalisation of the original mimetic gravity theory. We have proposed two simple toy models within the mimetic Horndeski class and showed that they possess interesting cosmological solutions. For instance, the simplest mimetic model is able to mimic the cosmological background evolution of a flat FLRW model with a barotropic perfect fluid with any constant equation of state (see also \cite{Lim:2010yk} for an earlier work). Actually by appropriately choosing the function $b(\varphi)$ in the transformation one can mimic almost any desired expansion history.

In the original mimetic model \cite{Chamseddine:2013kea} and its generalisation to include a potential \cite{Chamseddine:2014vna}, it was shown that the sound speed of scalar perturbations is exactly zero (independently of the desired expansion history) and consequently this model cannot describe a successful inflationary model because quantum fluctuations cannot be defined as usual. To circumvent this problem, it was proposed to introduce higher-derivatives terms in the action \cite{Chamseddine:2014vna}. In this way a non-zero sound speed can be generated. These higher-derivative terms help to suppress power for large momentum and it has been argued that this can be relevant for the small-scale  problems of cold dark matter \cite{Capela:2014xta}.

The main purposes of this work are to study linear scalar perturbations in mimetic Horndeski gravity and to determine the corresponding value of the sound speed. These results will determine the growth of structure in mimetic Horndeski models.

In the meantime, there have been many works studying different aspects of the original mimetic theory and generalisations. For example, the Hamiltonian analysis was performed in \cite{Malaeb:2014vua,Chaichian:2014qba}, cosmological perturbations were further analyzed in \cite{Matsumoto:2015wja}, extensions to $f(R)$ type models were presented in \cite{Leon:2014yua,Myrzakulov:2015qaa}, \cite{Momeni:2014qta} studied the energy conditions and a generalization,  a mimetic theory including a vector field was proposed in \cite{Barvinsky:2013mea}, cosmology in mimetic Galileon models studied in \cite{Haghani:2015iva,Rabochaya:2015haa}, and the imperfect fluid nature induced by higher-derivative terms was further discussed in \cite{Mirzagholi:2014ifa}.

This paper is organised as follows. In the next section, we introduce the model and some notation. We also show the general equations of motion of scalar-tensor mimetic gravity and discuss their independence. In section \ref{sec:LSP}, we discuss linear scalar perturbations of mimetic Horndeski in the Poisson gauge excluding other matter fields. We will compute the background equations of motion, then the first-order equations of motion for the Newtonian potential which we solve for the two toy models introduced in \cite{Arroja:2015wpa}. Section \ref{sec:SS} is devoted to the initial value formulation of the problem and to the discussion on the sound speed in general cosmological backgrounds. Section \ref{sec:CON} presents the conclusions of the paper. The paper has 4 appendices. In appendix \ref{BACK} we present the explicit expressions for the background equations of motion. Appendix \ref{FI} contains the expressions for the functions defined in the main text and that enter the first-order equations of motion. In appendix \ref{MATTEREOM}, we present the background and linear equations of motion for the mimetic Horndeski model including matter in the form of a fluid that may have anisotropic stress. Finally in appendix \ref{SSG3}, we compute the sound speed in a theory beyond mimetic Horndeski. We call this theory mimetic $G^3$ theory as it is the mimetic version of the so-called $G^3$ theory \cite{Gleyzes:2014dya,Gleyzes:2014qga}.

The reduced Planck mass is set to unity throughout the paper and we use the mostly plus metric signature.

\section{The model and notation\label{sec:MOD}}

In this section we will start by introducing a very general scalar-tensor theory of gravity including a term with a Lagrange multiplier following \cite{Arroja:2015wpa}. We will mostly follow notation of~\cite{Arroja:2015wpa} which we briefly summarise in this section too. There, it was shown that mimetic gravity has a dual formulation. One formulation is via a (non-invertible) disformal transformation as discussed in the introduction and the other formulation, that we will follow in this paper, is by using a Lagrange multiplier to impose the so-called mimetic constraint. In the following sections, where we will present explicit results for linear cosmological perturbations, we will restrict the very general mimetic scalar-tensor theory to the mimetic Horndeski theory. Horndeski's theory \cite{Horndeski:1974wa} is the most general 4D covariant theory of scalar-tensor gravity that is derived from an action and gives rise to second-order equations of motion (in all gauges and in any background) for both the metric and the scalar field. This useful property guarantees that the mimetic theory is free from higher-derivative ghosts because, as shown in \cite{Arroja:2015wpa}, if the original theory is free from these ghosts then also the mimetic theory that it originates is free from them.

Let us start with the very general action for scalar-tensor mimetic gravity as
\begin{eqnarray}
S&=&\int d^4x\sqrt{-g}\mathcal{L}[g_{\mu\nu},\partial_{\lambda_1}g_{\mu\nu},\ldots,\partial_{\lambda_1}\ldots\partial_{\lambda_p}g_{\mu\nu},\varphi,\partial_{\lambda_1}\varphi,\ldots,\partial_{\lambda_1}\ldots\partial_{\lambda_q}\varphi]+S_m[g_{\mu\nu},\phi_m]\nonumber\\
&+&\int d^4x\sqrt{-g}\lambda\left(b(\varphi)g^{\mu\nu}\partial_\mu\varphi\partial_\nu\varphi-1\right)
,\label{generalaction}
\end{eqnarray}
where $\lambda$ is a Lagrange multiplier and $\varphi$ is the mimetic scalar field. $b(\varphi)$ is a known potential function, $p,q\geq 2$ are integers. $S_m$ denotes the action for some matter field $\phi_m$ which we assume that is coupled with $g_{\mu\nu}$ only.
The equations of motion that result from varying the action with respect to $\lambda$, $\varphi$, $g_{\mu\nu}$ and $\phi_m$ are respectively (after some simplification)
\begin{eqnarray}
&&
b(\varphi)g^{\mu\nu}\partial_\mu\varphi\partial_\nu\varphi-1=0,\label{normalization}\\
&&
\Omega_\varphi+\sqrt{-g}\frac{\lambda}{b(\varphi)}\frac{db(\varphi)}{d\varphi}-2\partial_\mu\left(\sqrt{-g}\lambda b(\varphi)g^{\mu\nu}\partial_\nu\varphi\right)=0,\\
&&
E^{\mu\nu}+T^{\mu\nu}-2\lambda b(\varphi)\partial^\mu\varphi\partial^\nu\varphi=0,\label{mEE}\\
&&
\Omega_m=0,
\end{eqnarray}
where
\begin{eqnarray}
\Omega_\varphi&=&\frac{\delta\left(\sqrt{-g}\mathcal{L}\right)}{\delta\varphi}=\frac{\partial(\sqrt{-g}\mathcal{L})}{\partial\varphi}+\sum_{h=1}^q(-1)^h\frac{d}{dx^{\lambda_1}}\ldots\frac{d}{dx^{\lambda_h}}\frac{\partial(\sqrt{-g}\mathcal{L})}{\partial\left(\partial_{\lambda_1}\ldots\partial_{\lambda_h}\varphi\right)},
\label{Omegaphi}
\\
E^{\mu\nu}&=&\frac{2}{\sqrt{-g}}\frac{\delta(\sqrt{-g}\mathcal{L})}{\delta g_{\mu\nu}}=\frac{2}{\sqrt{-g}}\left(\frac{\partial(\sqrt{-g}\mathcal{L})}{\partial g_{\mu\nu}}+\sum_{h=1}^p(-1)^h\frac{d}{dx^{\lambda_1}}\ldots\frac{d}{dx^{\lambda_h}}\frac{\partial(\sqrt{-g}\mathcal{L})}{\partial\left(\partial_{\lambda_1}\ldots\partial_{\lambda_h}g_{\mu\nu}\right)}\right),
\\
T^{\mu\nu}&=&\frac{2}{\sqrt{-g}}\frac{\delta (\sqrt{-g}\mathcal{L}_m)}{\delta g_{\mu\nu}}, \,
\Omega_m=\frac{\delta (\sqrt{-g}\mathcal{L}_m)}{\delta \phi_m},
\,\mathrm{where}\,\, S_m[g_{\mu\nu},\phi_m]=\int d^4x\sqrt{-g}\mathcal{L}_m[g_{\mu\nu},\phi_m],
\end{eqnarray}
where $\mathcal{L}_m$ is the matter Lagrangian density and $T^{\mu\nu}$ denotes its energy-momentum tensor.
Taking the trace of Eq. (\ref{mEE}) and using Eq. (\ref{normalization}), one obtains
\begin{equation}
2\lambda=E+T,\label{sollambda}
\end{equation}
where $E=g_{\mu\nu}E^{\mu\nu}$ and $T=g_{\mu\nu}T^{\mu\nu}$. One can see that the Lagrange multiplier is given by the traces $E$ and $T$ and this can be used to eliminate $\lambda$ from the equations of motion to obtain
\begin{eqnarray}
&&
b(\varphi)g^{\mu\nu}\partial_\mu\varphi\partial_\nu\varphi-1=0,\label{normalization2}\\
&&
\nabla_\mu\left[(E+T)b(\varphi)\partial^\mu\varphi\right]-\frac{\Omega_\varphi}{\sqrt{-g}}=\frac{E+T}{2}\frac{1}{b(\varphi)}\frac{db(\varphi)}{d\varphi},\label{mKG}\\
&&
E^{\mu\nu}+T^{\mu\nu}=(E+T)b(\varphi)\partial^\mu\varphi\partial^\nu\varphi,\label{mEE2}\\
&&
\Omega_m=0\label{mMF}.
\end{eqnarray}
The previous set of equations are the equations of motion for the theory (\ref{generalaction}). However not all the equations in the set are independent from each other. As shown in \cite{Arroja:2015wpa}, Eq. (\ref{mKG}) can be derived from the other equations. Also as we will now show, the $0-0$ component of Eqs. (\ref{mEE2}) can be derived from Eq. (\ref{normalization}) and the remaining components of Eqs. (\ref{mEE2}).

Let us start with the constraint equation
\begin{eqnarray}
b(\varphi)g^{\mu\nu}\partial_\mu\varphi\partial_\nu\varphi&=&b(\varphi)g^{00}(\varphi')^2+2b(\varphi)g^{0i}(\varphi')\partial_i\varphi+b(\varphi)g^{ij}\partial_i\varphi\partial_j\varphi=1,
\end{eqnarray}
where $'$ denotes the derivative with respect to the time coordinate (which in the next section we choose to be conformal time).
Multiply both sides of the previous equation by $E+T$ to obtain
\begin{eqnarray}
&&(E+T)b(\varphi)g^{00}(\varphi')^2+2(E+T)b(\varphi)g^{0i}(\varphi')\partial_i\varphi+(E+T)b(\varphi)g^{ij}\partial_i\varphi\partial_j\varphi
\nonumber\\
&&\qquad\qquad\qquad
=g^{00}(E_{00}+T_{00})+2g^{0i}(E_{0i}+T_{0i})+g^{ij}(E_{ij}+T_{ij}),\label{eqaux}
\end{eqnarray}
where Latin indexes run from one to three only.
By using the other components of Eqs. (\ref{mEE2}), i.e.
\begin{equation}
E_{ij}+T_{ij}=(E+T)b(\varphi)\partial_i\varphi\partial_j\varphi,\quad
E_{0i}+T_{0i}=(E+T)b(\varphi)\varphi'\partial_i\varphi,\label{eomaux}
\end{equation}
one can show that Eq. (\ref{eqaux}) simplifies to
\begin{eqnarray}
(E+T)b(\varphi)g^{00}(\varphi')^2=g^{00}(E_{00}+T_{00}).
\end{eqnarray}
Because $g^{00}\neq0$ we have the desired result that Eqs. (\ref{eomaux}) together with the constraint equation imply
\begin{equation}
E_{00}+T_{00}=(E+T)b(\varphi)(\varphi')^2.
\end{equation}
This is a non-perturbative result and it will be important when counting the number of perturbation variables and their equations in the next section.

In the remainder of the paper, for the reasons previously mentioned, we will only consider a particular subset of theories of the form (\ref{generalaction}) known as mimetic Horndeski theory. The action of mimetic Horndeski gravity is defined by Eq. (\ref{generalaction}) where the Lagrangian density $\mathcal{L}$ is given by Horndeski's Lagrangian density $\mathcal{L}_H$ as
\begin{equation}
S_H=\int d^4x\sqrt{-g}\mathcal{L}_H=\int d^4x\sqrt{-g}\sum_{n=0}^{3}\mathcal{L}_{n},
\label{action}
\end{equation}
where
\begin{eqnarray}
\mathcal{L}_0 & = & K\left(X,\varphi\right),\\
\mathcal{L}_1 & = & -G_3\left(X,\varphi\right)\Box\varphi,\\
\mathcal{L}_2 & = & G_{4,X}\left(X,\varphi\right)\left[\left(\Box\varphi\right)^{2}-\left(\nabla_{\mu}\nabla_{\nu}\varphi\right)^{2}\right]
+R\,G_4\left(X,\varphi\right) ,\\
\mathcal{L}_3 & = & -\frac{1}{6}G_{5,X}\left(X,\varphi\right)\left[\left(\Box\varphi\right)^{3}-3\Box\varphi\left(\nabla_{\mu}\nabla_{\nu}\varphi\right)^{2}
+2\left(\nabla_{\mu}\nabla_{\nu}\varphi\right)^{3}\right]+G_{\mu\nu}\nabla^{\mu}\nabla^{\nu}\varphi\,
G_5\left(X,\varphi\right),
\end{eqnarray}
and $X=-1/2\nabla_\mu\varphi\nabla^\mu\varphi$, $(\nabla_\mu\nabla_\nu\varphi)^2=\nabla_\mu\nabla_\nu\varphi\nabla^\mu\nabla^\nu\varphi$ and $(\nabla_\mu\nabla_\nu\varphi)^3=\nabla_\mu\nabla_\nu\varphi\nabla^\mu\nabla^\rho\varphi\nabla^\nu\nabla_\rho\varphi$.
The subscript $,X$ denotes derivative with respect to $X$ and in the following derivative with respect to $\varphi$ will be denoted by a subscript $,\varphi$. The functions $K,\,G_3,\,G_4,\,G_5$ of two variables, $X$ and $\varphi$, define a particular (mimetic) Horndeski theory.

\section{Linear scalar perturbations \label{sec:LSP}}

This section is devoted to the study of cosmological linear scalar perturbations in the mimetic Horndeski gravity. Here we will assume that there is no matter in the model, i.e. $S_m=0$. We expect this to be a good approximation during the time when the effective energy density of the mimetic scalar field is much larger than the other usual components of the total energy density like radiation or cold dark matter. In appendix \ref{MATTEREOM} we present the equations of motion of the mimetic Horndeski model including a matter source in the form of a fluid which may have anisotropic stress as it would be the case for free-streaming neutrinos. Before that, in the next subsection we will present well-known (see for instance \cite{Kobayashi:2011nu,DeFelice:2011hq}) results for linear scalar perturbations in Horndeski gravity, as a warm up.

We will work in the Poisson gauge. Because we are only interested in scalar perturbations, we will neglect vector and tensor perturbations. At linear order and in the flat FLRW background that we will assume, these different type of perturbations are all decoupled.

The metric is perturbed as
\begin{equation}
g_{00}=-a^2(\tau)\left(1+2\Phi\right),\quad g_{0i}=0, \quad g_{ij}=a^2(\tau)\left(1-2\Psi\right)\delta_{ij},
\end{equation}
where $a$ is the FLRW scale factor that depends on the conformal time $\tau$, $\Phi$ denotes the generalised Newtonian (Bardeen) potential and $\Psi$ the curvature perturbation. The inverse metric is
\begin{equation}
g^{00}=-a^{-2}(\tau)\left(1-2\Phi\right),\quad g^{0i}=0, \quad g^{ij}=a^{-2}(\tau)\left(1+2\Psi\right)\delta^{ij}.
\end{equation}
The scalar field is expanded as $\varphi(\tau,\mathbf{x})=\varphi_0(\tau)+\delta\varphi(\tau,\mathbf{x})$, where $\varphi_0$ denotes the background field value and $\delta\varphi$ is the field perturbation.

\subsection{Linear scalar perturbations in Horndeski\label{subsec:PERTHORNDESKI}}

We will study linear perturbations of Horndeski gravity only in this subsection. The theory is defined by the action (\ref{action}). The tensor $E^{\mu\nu}$ introduced in the previous section will be the same for both Horndeski and mimetic Horndeski gravity as it is clear from its definition.

Because we assume that there are no matter sources and the equation of motion for $\varphi$ is not independent from the metric equations of motion as it is well-known\footnote{This well-known fact can be simply understood to be a consequence of Horndeski's identity \cite{Horndeski:1974wa} (see also references therein), i.e., $\sqrt{-g}\nabla_\mu E^{\mu\nu}=\Omega_\varphi\nabla^\nu\varphi$. For a general scalar-tensor theory defined by the first line of Eq. (\ref{generalaction}), which includes Horndeski's theory as a particular case, the equation of motion for the scalar field is $\Omega_\varphi=0$, which implies, by using the previous identity, $\nabla_\mu E^{\mu\nu}=0$. The previous equation is the generalization of the usual equation for the conservation of the energy-momentum tensor. Eq. (\ref{Horndeskieom}) automatically implies that the equation of motion for the scalar field is satisfied.}, the equations of motion are simply
\begin{equation}
E_{\mu\nu}=0.\label{Horndeskieom}
\end{equation}

At the background level they reduce to $E_{\mu\nu}^{(0)}=0$, where the superscript $(0)$ denotes background quantities and the explicit expressions for $E_{\mu\nu}^{(0)}$ in terms of the Horndeski functions and their derivatives can be found in appendix \ref{BACK}.

At first order (denoted by the superscript $(1)$) the tensor $E_{\mu\nu}$ can be written as
\begin{eqnarray}
E_{00}^{(1)}&=&f_1\Psi'+f_2\delta\varphi'+f_3\Phi+f_4\delta\varphi+f_5\partial^2\Psi+f_6\partial^2\delta\varphi,\label{E00Horndeski}
\\
E_{ij}^{(1)}&=&\partial_i\partial_j\left(f_7\Psi+f_8\delta\varphi+f_9\Phi\right)
+\delta_{ij}\Big(-f_7\partial^2\Psi-f_8\partial^2\delta\varphi-f_9\partial^2\Phi
\nonumber\\
&&+f_{10}\Psi''+f_{11}\delta\varphi''+f_{12}\Psi'+f_{13}\delta\varphi'+f_{14}\Phi'+f_{15}\Psi+f_{16}\delta\varphi+f_{17}\Phi\Big),\label{EijHorndeski}
\\
E_{0i}^{(1)}&=&\partial_i\left(f_{18}\Psi'+f_{19}\delta\varphi'+f_{20}\delta\varphi+f_{21}\Phi\right),\label{E0iHorndeski}
\end{eqnarray}
where $'$ denotes derivative with respect to conformal time, the functions $f_i$, $i=1,...,21$ are linear functions of $K, G_3, G_4, G_5$ and their derivatives evaluated on the background, therefore the $f_i$ are functions of time only. Their explicit expressions are given in appendix \ref{FI}. These functions are not all independent from each other and they obey certain relations also given in appendix \ref{FI}.

For a standard kinetic term scalar field coupled to Einstein gravity, at first order, it is well known that Eqs. (\ref{Horndeskieom}) are not all independent. By taking the time derivative of the $E_{0i}^{(1)}=0$ equation and using it again together with the background equations it is possible to obtain the evolution part of the $E_{ij}^{(1)}=0$ equation (which corresponds to the second line of Eq.~(\ref{EijHorndeski})). In Horndeski theory, something similar should also happen as we will discuss below. By taking the traceless part of $E_{ij}^{(1)}=0$ one can see that the first line of $E_{ij}^{(1)}$ vanishes. In other words the traceless part of $E_{ij}^{(1)}=0$ implies that
\begin{equation}
f_7\Psi+f_8\delta\varphi+f_9\Phi=0.
\end{equation}
The physical implications of the previous equation are that the anisotropic stress is in general not zero and also that at least one of the fields is not a new dynamical degree of freedom. This equation will also be valid in the mimetic Horndeski case.

Let us now count the number of variables and equations of motion. We have three variables, $\delta\varphi$, $\Phi$ and $\Psi$ to be determined by the equations of motion, $E_{\mu\nu}^{(1)}=0$ which can be written as
\begin{eqnarray}
f_1\Psi'+f_2\delta\varphi'+f_3\Phi+f_4\delta\varphi+f_5\partial^2\Psi+f_6\partial^2\delta\varphi=0,\label{EOM00Horndeski}
\\
f_7\Psi+f_8\delta\varphi+f_9\Phi=0,\label{EOMijHorndeskiTraceless}
\\
f_{10}\Psi''+f_{11}\delta\varphi''+f_{12}\Psi'+f_{13}\delta\varphi'+f_{14}\Phi'+f_{15}\Psi+f_{16}\delta\varphi+f_{17}\Phi=0,\label{EOMijHorndeskiTrace}
\\
f_{18}\Psi'+f_{19}\delta\varphi'+f_{20}\delta\varphi+f_{21}\Phi=0.\label{EOM0iHorndeski}
\end{eqnarray}
Naively one has three variables and four equations, however, one can show that there are only three independent equations, as expected. Eq. (\ref{EOMijHorndeskiTrace}) can be derived from Eqs. (\ref{EOMijHorndeskiTraceless}) and (\ref{EOM0iHorndeski}) and by using some of the identities in appendix \ref{FI} \footnote{One can follow a brute force procedure: use Eq. (\ref{EOMijHorndeskiTraceless}) to write $\Psi$ in terms of the other variables. Use Eq. (\ref{EOM0iHorndeski}) and solve it for $\Phi'$. Take the time derivative of Eq. (\ref{EOM0iHorndeski}) and sum to it a term $T\times\Phi'$ and then subtract from it the same term $T\times\Phi'$ but where now $\Phi'$ is replaced with the previously found expression and $T=4\mathcal{H}f_9^2/f_7$, where $\mathcal{H}$ is defined as $\mathcal{H}=a'/a$. The equation obtained differs from Eq. (\ref{EOMijHorndeskiTrace}) in terms proportional to $\Phi$ and $\delta\varphi$ only. However, by using the identities in appendix \ref{FI} one can show that actually the coefficients of $\Phi$ and $\delta\varphi$ terms are exactly the ones in Eq. (\ref{EOMijHorndeskiTrace}).}.

\subsection{Linear scalar perturbations in mimetic Horndeski\label{subsec:PERTMIMETICHORNDESKI}}

We now turn to the main goal of this paper, i.e., to study linear scalar perturbations in mimetic Horndeski gravity. As we have explained in Sec.~\ref{sec:MOD}
the independent equations of motion for the model reduce to (assuming that there is no matter; see appendix \ref{MATTEREOM} for the case when matter is present)

\begin{eqnarray}
b(\varphi)g^{\mu\nu}\partial_\mu\varphi\partial_\nu\varphi-1=0,\qquad
E_{\mu i}=Eb(\varphi)\partial_\mu\varphi\partial_i\varphi.
\end{eqnarray}

At zeroth order on a flat FLRW background, they simplify to
\begin{eqnarray}
-a^{-2}b_0(\varphi_0')^2&=&1,\label{constraint0th}\\
E_{ij}^{(0)}&=&0,
\end{eqnarray}
where $b_0$ denotes $b(\varphi_0)$.

The first-order equations of motion are
\begin{eqnarray}
2b_0\delta\varphi'+\varphi_0'b_{,\varphi}\delta\varphi-2b_0\varphi_0'\Phi&=&0,\label{constraint1st}\\
E_{ij}^{(1)}&=&0,\label{ijequation}\\
E_{0i}^{(1)}&=&E^{(0)}b_0\varphi_0'\partial_i\delta\varphi,\label{momentumequation}
\end{eqnarray}
where $E^{(0)}$ denotes the zeroth-order trace of $E_{\mu\nu}$, $b_{,\varphi}=b_{,\varphi}(\varphi_0)$ and the subscript ${,\varphi}$ denotes derivative with respect to the field $\varphi$. As mentioned in the previous subsection the $E_{\mu\nu}$ tensor is equal to the one defined for the Horndeski's theory whose explicit expressions are given by Eqs. (\ref{E00Horndeski})-(\ref{E0iHorndeski}).
Eq. (\ref{ijequation}) implies
\begin{eqnarray}
&&f_7\Psi+f_8\delta\varphi+f_9\Phi=0,\label{Psiequation}
\\
&&f_{10}\Psi''+f_{11}\delta\varphi''+f_{12}\Psi'+f_{13}\delta\varphi'+f_{14}\Phi'+f_{15}\Psi+f_{16}\delta\varphi+f_{17}\Phi=0,\label{evolutionequation}
\end{eqnarray}
and Eq. (\ref{momentumequation}) implies
\begin{equation}
f_{18}\Psi'+f_{19}\delta\varphi'+\left(f_{20}+\frac{a^2E^{(0)}}{\varphi_0'}\right)\delta\varphi+f_{21}\Phi=0,\label{momentumconstraint}
\end{equation}
where we have used the zeroth-order constraint. As before, Eq. (\ref{Psiequation}) can be used for example to eliminate $\Psi$ from all the equations in favor of the pair $\delta\varphi,\Phi$. In other words, $\Psi$ is not a new degree of freedom with respect to the pair $\delta\varphi,\Phi$. One can also see that in general $\Psi\neq\Phi$ for mimetic Horndeski (i.e., there is some non-zero effective anisotropic stress).

It is important to note that because Horndeski's theory is form-invariant under a field redefinition one can without loss of generality set $b(\varphi)=-1$ in mimetic Horndeski. In that case, $b_{,\varphi}=0$ and then the first-order constraint implies $\Phi=\frac{\delta\varphi'}{\varphi_0'}$.

At this point, we have four equations of motion, Eqs. (\ref{constraint1st}), (\ref{Psiequation}), (\ref{evolutionequation}) and (\ref{momentumconstraint}), and only three variables, $\Psi$, $\Phi$ and $\delta\varphi$. However, following a similar procedure to the one in the previous subsection one can show that (\ref{evolutionequation}) can be derived from the other two equations, i.e., Eqs. (\ref{Psiequation}) and (\ref{momentumconstraint}), where this time to complete the proof one also needs to use Eq. (\ref{constraint1st}).
In summary, the independent first-order equations of motion for the mimetic Horndeski model that we will use from now on are
\begin{eqnarray}
&&2b_0\delta\varphi'+\varphi_0'b_{,\varphi}\delta\varphi-2b_0\varphi_0'\Phi=0,\label{seteom1}\\
&&f_7\Psi+f_8\delta\varphi+f_9\Phi=0,\label{seteom2}
\\
&&f_{18}\Psi'+f_{19}\delta\varphi'+\left(f_{20}+\frac{a^2E^{(0)}}{\varphi_0'}\right)\delta\varphi+f_{21}\Phi=0. \label{seteom3}
\end{eqnarray}
Because in this system of equations there are no spatial derivatives one can anticipate that the sound speed for the dynamical scalar degree of freedom will be exactly zero.

Indeed, from the previous three equations one can find an evolution equation for the Newtonian potential $\Phi$ as
\begin{eqnarray}
\Phi''+\left(\frac{B_2}{B_3}+\left(\ln\frac{B_3}{B_1}\right)'+\mathcal{H}-\frac{\varphi_0''}{\varphi_0'}\right)\Phi'+\left(\frac{B_1}{B_3}\varphi_0'+\frac{B_1}{B_3}\left(\frac{B_2}{B_1}\right)'+\frac{B_2}{B_3}\left(\mathcal{H}-\frac{\varphi_0''}{\varphi_0'}\right)\right)\Phi=0,
\label{eomPhi}
\end{eqnarray}
where the $B_i$ functions are defined as
\begin{eqnarray}
B_1&=&f_{20}+\frac{f_{10}f_8f_7'}{f_7^2}
+f_{11}\left(-\mathcal{H}+\frac{\varphi_0''}{\varphi_0'}\right)-\frac{f_{10}}{f_7}\left(f_8'+f_8\left(-\mathcal{H}+\frac{\varphi_0''}{\varphi_0'}\right)\right)+a^2\frac{E^{(0)}}{\varphi_0'},
\\
B_2&=&f_{14}+\frac{f_{10} f_9 f_7'}{f_7^2} +
 f_{11}\varphi_0' - \frac{f_{10}(f_9'+f_8\varphi_0')}{f_7},
\\
B_3&=&2\frac{f_9^2}{f_7}.
\end{eqnarray}
There is no spatial Laplacian term so this means that the sound speed of the perturbations is exactly zero as anticipated. In appendix \ref{SSG3}, we show that the conclusion that the sound speed is exactly zero also applies to a scalar-tensor theory more general than mimetic Horndeski, the one which is built starting from $G^3$ theories \cite{Gleyzes:2014dya,Gleyzes:2014qga}. Notice also that in Eq.~(\ref{eomPhi}) there is no source term on the right-hand side, which is usually associated to the presence of entropy perturbation modes,
see, e.g.,~\cite{Mukhanov:1990me}. This will be also confirmed below when deriving the equation of motion for the comoving curvature perturbation.
After solving Eq. (\ref{eomPhi}) for $\Phi$ one can use the constraint equation, Eq. (\ref{seteom1}), for a given function $b(\varphi)$, to solve for $\delta\varphi$. Finally to find $\Psi$ one can solve
\begin{equation}
\delta\varphi=\left(-2\mathcal{H}f_8+f_{11}\varphi_0'\frac{b_{,\varphi}}{2b_0}-f_{20}-\frac{a^2E^{(0)}}{\varphi_0'}\right)^{-1}\left(f_{10}\Psi'+2\mathcal{H}f_7\Psi\right).
\end{equation}
It is convenient to introduce a new variable, the comoving curvature perturbation $\zeta$ (the comoving gauge is defined by $\delta\varphi=0$ and in that gauge $\zeta$ is related (to first order) to the 3D curvature as $R^{(3)}=-4a^{-2}\partial^2\zeta$), which is defined as
\begin{equation}
-\zeta=\Psi+\frac{\mathcal{H}}{\varphi_0'}\delta\varphi.
\end{equation}
One can show that the set of equations of motion of the model, Eqs. (\ref{seteom1})-(\ref{seteom3}), is equivalent to (using the background equations of motion)
\begin{eqnarray}
&&2b_0\delta\varphi'+\varphi_0'b_{,\varphi}\delta\varphi-2b_0\varphi_0'\Phi=0,\label{seteomccp1}\\
&&-f_7\zeta+\left(f_8-\frac{\mathcal{H}}{\varphi_0'}f_7\right)\delta\varphi+f_9\Phi=0,\label{seteomccp2}
\\
&&\zeta'=0. \label{seteomccp3}
\end{eqnarray}
Note that the comoving curvature perturbation $\zeta$ has a first-order equation of motion with solution $\zeta=\textrm{constant}$ on all scales (and vanishing
intrinsic entropy perturbations, see, e.g.~\cite{Bartolo:2003ad}).

For the particular case when $G_4(X,\varphi)=1/2$ and $G_5(X,\varphi)=0$ (and the other functions of the Horndeski theory, i.e. $K$ and $G_3$, are kept general) we were able to simplify the evolution equation, Eq. (\ref{eomPhi}), to obtain
\begin{equation}
\Phi''+\Phi'\left(3\mathcal{H}+\tilde\Gamma\right)+\Phi\left(\mathcal{H}^2+2\mathcal{H}'+\tilde\Gamma\mathcal{H}\right)=0,\label{Phieom}
\end{equation}
where the variable $\tilde\Gamma$ is defined as
\begin{equation}
\tilde\Gamma=\frac{-\mathcal{H}''+\mathcal{H}\mathcal{H}'+\mathcal{H}^3}{\mathcal{H}'-\mathcal{H}^2}.
\end{equation}
The quantity $\tilde\Gamma$ can be seen as a correction to the perturbation equation of standard pressureless dust that arises in these mimetic models. This equation was first derived and solved in \cite{Lim:2010yk} for the case when the function $G_3(X,\varphi)$ was zero. What we found is that this equation is still valid even if $G_3(X,\varphi)\neq0$.
Let us note that the particular mimetic models for which Eq. (\ref{Phieom}) is valid include the two models studied in \cite{Arroja:2015wpa} that showed very interesting cosmological behaviour (e.g. they can reproduce exactly the $\Lambda$CDM background expansion).
It is important to note that the quantity $\tilde\Gamma$ is written in a geometrical way and that it exactly vanishes for a $\Lambda$CDM expansion history. Indeed, the differential equation $\tilde\Gamma=0$ has the three solutions: $a(t)\propto\exp^{\mathcal{C}t}$, $a(t)\propto t^{2/3}$ and $a(t)\propto\sinh^\frac{2}{3}(\mathcal{C} t)$. In the limit of a $\Lambda$CDM background expansion history, corresponding to the latter solution, the perturbations in these particular mimetic Horndeski models will behave exactly in the same way as the perturbations in a $\Lambda$CDM universe.
For the particular case when $G_4(X,\varphi)=1/2$ and $G_5(X,\varphi)=0$, the relation between $\zeta$ and $\Phi$ is
\begin{equation}
\zeta=-\frac{2\mathcal{H}^2-\mathcal{H}'}{\mathcal{H}^2-\mathcal{H}'}\Phi-\frac{\mathcal{H}}{\mathcal{H}^2-\mathcal{H}'}\Phi',
\end{equation}
and $\Psi=\Phi$.

To find the solutions of Eq. (\ref{Phieom}) let us use the results of \cite{Lim:2010yk}.
By using the number of e-folds, $N=\ln a$, as the time variable and using a new variable as
\begin{equation}
Q=\sqrt{\frac{a}{-H_{,N}}}\Phi,
\end{equation}
where the subscript $,N$ denotes derivative with respect to $N$, the Hubble rate is defined as $H=da/(adt)$, where $t$ denotes cosmic time, one can write the evolution equation, Eq. (\ref{Phieom}), as
\begin{equation}
Q_{,NN}-\frac{\Theta_{,NN}}{\Theta}Q=0,\label{Qequation}
\end{equation}
where the variable $\Theta$ is defined as
\begin{equation}
\Theta=\frac{H}{\sqrt{-aH_{,N}}}.
\end{equation}
It is immediate to show that $Q\propto\Theta$ is a solution of Eq. (\ref{Qequation}). The other solution can be found using the Wronskian method and is $Q\propto\sqrt{\frac{a}{-H_{,N}}}\left(1-\frac{H}{a}\int\frac{da}{H}\right)$. This implies that the two solutions for the Newtonian potential are
\begin{eqnarray}
\Phi_1=\frac{H}{a},\qquad \Phi_2=1-\frac{H}{a}\int\frac{da}{H}.
\end{eqnarray}
The general solution for $\Phi$ is a linear combination of $\Phi_1$ and $\Phi_2$, with $C_1(\mathbf{x})$ and $C_2(\mathbf{x})$ being the integration constants, as
\begin{equation}
\Phi(\tau,\mathbf{x})=C_1(\mathbf{x})+\frac{H}{a}C_2(\mathbf{x})-C_1(\mathbf{x})\frac{H}{a}\int\frac{da}{H}.
\end{equation}
The same solutions were found in \cite{Lim:2010yk} for a model with $G_4(X,\varphi)=1/2$ and $G_3(X,\varphi)=G_5(X,\varphi)=0$. Here we show that the form of the solutions is the same even if $G_3(X,\varphi)\neq0$. The solution $\Phi_1$ corresponds to $\zeta=0$ while the solution $\Phi_2$ corresponds to $\zeta=\textrm{constant}\neq0$.
If the scale factor is $a\propto t^\frac{2}{3(1+w)}$, where $w$ is the equation of state, then $H/a\propto a^\frac{-5-3w}{2}$ which decays for an expanding universe if $w>-5/3$ and therefore $\Phi_1$ is the decaying mode in that case ($\Phi_2$ is constant).

\section{The Cauchy problem and the sound speed\label{sec:SS}}

In this section, we will follow the method of \cite{Lim:2010yk} to show that, without assuming any background but for a non-dynamical metric, the sound speed is exactly zero in a general mimetic Horndeski theory (without additional matter fields). We will assume that the four-velocity is time-like because we have cosmology applications in mind.

The constraint equation, Eq. (\ref{normalization2}), which is the equation of motion for the Lagrange multiplier, implies that
\begin{equation}
b(\varphi)X=-\frac{1}{2},\label{simcons}
\end{equation}
where we assume that $b(\varphi)<0$. For a mimetic Horndeski theory, one can redefine the scalar field as to absorb  $b(\varphi)$, or in other words, one can choose  $b(\varphi)=-1$ without losing generality \cite{Arroja:2015wpa}. From now on in this section we will set  $b(\varphi)=-1$. The constraint then implies $X=1/2$.
One can define the four-velocity as
\begin{equation}
u_\nu=\frac{\nabla_\nu\varphi}{\sqrt{2X}}=\nabla_\nu\varphi,
\end{equation}
which satisfies the constraint $u_\nu u^\nu=-1$ and in the last equality we have used the constraint $X=1/2$. In this section we will use the notation $\dot{(\,\,)}=u^\nu\nabla_\nu(\,\,)$ to denote the derivative along $u^\nu$.
It is easy to see that the four-acceleration, $a^\nu$, is always zero, i.e.
\begin{equation}
a^\nu=\dot u^\nu=u^\mu\nabla_\mu u^\nu=0,\label{acceleration}
\end{equation}
because of the constraint equation, Eq. (\ref{simcons}). This means the flow is always geodesic.
The constraint equation becomes very simple, it reads
\begin{equation}
\dot\varphi=-1.\label{fieldlambdaeq}
\end{equation}

Let us decompose the covariant derivative of $u_\nu$ in its symmetric and skew-symmetric parts as
\begin{equation}
\nabla_\mu u_\nu=\theta_{\mu\nu}+\omega_{\mu\nu},
\end{equation}
where
\begin{equation}
\theta_{\mu\nu}=\nabla_{(\mu}u_{\nu)}, \quad \omega_{\mu\nu}=\nabla_{[\mu}u_{\nu]},\quad \theta_{\mu\nu}=\theta_{(\mu\nu)}=\frac{1}{2}\left(\theta_{\mu\nu}+\theta_{\nu\mu}\right),\quad \omega_{\mu\nu}=\omega_{[\mu\nu]}=\frac{1}{2}\left(\omega_{\mu\nu}-\omega_{\nu\mu}\right).
\end{equation}
The tensor $\theta_{\mu\nu}$ is called the expansion tensor and the tensor $\omega_{\mu\nu}$ is called the vorticity tensor. These satisfy $\theta_{\mu\nu}u^\mu=\omega_{\mu\nu}u^\mu=0$. In the present case of a mimetic scalar field, the vorticity tensor is zero because $\nabla_\mu u_\nu=\nabla_\nu u_\mu$. Let us further decompose the expansion tensor in its trace and trace-free parts as
\begin{equation}
\theta_{\mu\nu}=\sigma_{\mu\nu}+\frac{1}{3} \theta h_{\mu\nu},
\end{equation}
where $h_{\mu\nu}$ is defined as $h_{\mu\nu}=g_{\mu\nu}+u_\mu u_\nu$. $\theta$ is called the volume expansion and $\sigma_{\mu\nu}$ is called the shear tensor. These satisfy $\sigma_{\mu\nu}u^\mu=\sigma_\nu^\nu=0$ where $\sigma_\nu^\nu=g^{\mu\nu}\sigma_{\mu\nu}$ and $\sigma_{(\mu\nu)}=\sigma_{\mu\nu}$.
One can then find
\begin{equation}
\theta=g^{\mu\nu}\theta_{\mu\nu}=\nabla_\nu u^\nu, \qquad \sigma_{\mu\nu}=\nabla_\mu u_\nu-\frac{1}{3}\theta\left(g_{\mu\nu}+u_\mu u_\nu\right).
\end{equation}
$\theta$ satisfies the well-known Landau-Raychaudhuri equation
\begin{equation}
\dot\theta=-\sigma_{\mu\nu}\sigma^{\mu\nu}-\frac{\theta^2}{3}-R_{\mu\nu}u^\mu u^\nu.\label{LReq}
\end{equation}
The evolution equation for the shear tensor is
\begin{equation}
\dot\sigma_{\mu\nu}=-\sigma^\lambda_\mu\sigma_{\lambda\nu}-\frac{2}{3}\theta\sigma_{\mu\nu}-\frac{1}{3}h_{\mu\nu}\left(\dot\theta+\frac{1}{3}\theta^2\right)-R_{\alpha\lambda\beta\rho} u^\lambda u^\rho.\label{sheareq}
\end{equation}
In deriving the previous two equations we have used several times Eq. (\ref{acceleration}), $(\nabla_\alpha\nabla_\beta-\nabla_\beta\nabla_\alpha)V^\mu=R^\mu_{~\nu\alpha\beta}V^\nu$ for a vector $V^\mu$, $R_{\mu\nu}=R^\lambda_{~\mu\lambda\nu}$ and the general properties of the Riemann tensor $R_{\mu\nu\alpha\beta}$.

The equation of motion for the field $\varphi$ can be written as
\begin{equation}
2\dot\lambda+2\theta\lambda+\frac{\Omega_\varphi}{\sqrt{-g}}=0,\label{fieldvarphieq}
\end{equation}
where $\lambda$ is the Lagrange multiplier field introduced in Eq. (\ref{generalaction}), $\Omega_\varphi$, defined in Eq.~(\ref{Omegaphi}), can be written as $\Omega_\varphi=\sqrt{-g}\sum_{i=2}^5\left(P^{(i)}_\varphi-\nabla^\mu J^{(i)}_\mu\right)$ and the explicit (long) expressions of $P^{(i)}_\varphi$ and $J_\mu^{(i)}$ can be found in Appendix B of \cite{Kobayashi:2011nu}. For the mimetic Horndeski theory that we are interested in, those expressions can be written as
\begin{eqnarray}
P^{(2)}_\varphi-\nabla^\mu J^{(2)}_\mu&=&K_{,\varphi}-\left(K_{,X\varphi}-K_{,X}\theta\right),
\end{eqnarray}
\begin{eqnarray}
P^{(3)}_\varphi-\nabla^\mu J^{(3)}_\mu&=&-G_{3,\varphi\varphi}-\left[-2G_{3,\varphi\varphi}+\left(2G_{3,\varphi}-G_{3,X\varphi}\right)\theta+G_{3,X}\left(\theta^2+\dot\theta\right)\right],
\end{eqnarray}
\begin{eqnarray}
P^{(4)}_\varphi-\nabla^\mu J^{(4)}_\mu&=&G_{4,\varphi}R+G_{4,X\varphi}\left(\frac{2}{3}\theta^2-\sigma_{\mu\nu}\sigma^{\mu\nu}\right)
\nonumber\\
&&
-\Bigg[
G_{4,X}\left(-R\theta+2R_{\mu\nu}\left(\sigma^{\mu\nu}+\frac{1}{3}\theta h^{\mu\nu}\right)\right)
-G_{4,XX}\left(\frac{4}{3}\theta\dot\theta-2\sigma_{\mu\nu}\dot\sigma^{\mu\nu}+\left(\frac{2}{3}\theta^2-\sigma_{\mu\nu}\sigma^{\mu\nu}\right)\theta\right)
\nonumber\\
&&
\quad
+G_{4,X\varphi}\left(R+2u^\mu u^\nu R_{\mu\nu}-2\dot\theta-2\theta^2\right)+G_{4,XX\varphi}\left(\frac{2}{3}\theta^2-\sigma_{\mu\nu}\sigma^{\mu\nu}\right)
+2G_{4,X\varphi\varphi}\theta
\Bigg],
\end{eqnarray}
\begin{eqnarray}
P^{(5)}_\varphi-\nabla^\mu J^{(5)}_\mu&=&-G_{5,\varphi\varphi}u^\mu u^\nu G_{\mu\nu}-\frac{1}{6}G_{5,X\varphi}\left(2\sigma_{\mu\nu}\sigma^{\nu\lambda}\sigma^\mu_\lambda-\theta\sigma_{\mu\nu}\sigma^{\mu\nu}+\frac{2}{9}\theta^3\right)
\nonumber\\
&&
-\Bigg[
\left(\frac{1}{2}G_{5,X\varphi}\left(\frac{2}{3}\theta^2-\sigma_{\mu\nu}\sigma^{\mu\nu}\right)+\frac{1}{6}G_{5,XX}\left(2\sigma_{\mu\nu}\sigma^{\nu\lambda}\sigma^\mu_\lambda-\theta\sigma_{\mu\nu}\sigma^{\mu\nu}+\frac{2}{9}\theta^3\right)\right)^\centerdot
\nonumber\\
&&\quad
+\frac{1}{6}G_{5,XX}\theta\left(2\sigma_{\mu\nu}\sigma^{\nu\lambda}\sigma^\mu_\lambda-\theta\sigma_{\mu\nu}\sigma^{\mu\nu}+\frac{2}{9}\theta^3\right)
-G_{5,\varphi}\left(2R_{\mu\nu}\sigma^{\mu\nu}-\frac{1}{3}R\theta+\frac{2}{3}\theta u^\mu u^\mu R_{\mu\nu}\right)
\nonumber\\
&&\quad
-G_{5,X\varphi}\left(-R_{\mu\nu}\sigma^{\mu\nu}+\frac{1}{6}\theta R+\frac{1}{3}\theta u^\mu u^\nu R_{\mu\nu}-u^\mu u^\nu R_{\nu\lambda}\sigma^\lambda_\mu-u^\mu u^\nu \sigma^{\alpha\beta}R_{\alpha\mu\beta\nu}-\frac{1}{3}\theta^3+\frac{1}{2}\sigma_{\mu\nu}\sigma^{\mu\nu}\theta\right)
\nonumber\\
&&\quad
-G_{5,X}\Bigg(
R_{\mu\nu}\dot\sigma^{\mu\nu}-\frac{1}{6}\left(\dot\theta +\frac{1}{3}\theta^2\right)R+\frac{1}{3}R_{\mu\nu}\left(2\theta\sigma^{\mu\nu}+\left(\theta^2+\dot\theta\right)u^\mu u^\nu\right)
-R_{\mu\nu}\sigma^{\mu\lambda}\sigma_\lambda^\nu
\nonumber\\
&&\quad
-\sigma^{\alpha\beta}R_{\alpha\mu\beta\nu}\left(\sigma^{\mu\nu}+\frac{2}{3}\theta u^\mu u^\nu\right)-R_{\mu\nu}R_\lambda^\mu u^\nu u^\lambda-u^\nu u_\lambda R_{\alpha\mu\beta\nu}R^{\mu\beta\alpha\lambda}
\Bigg)-G_{5,\varphi\varphi}\left(2u^\mu u^\nu R_{\mu\nu}+R\right)
\Bigg],\nonumber\\
\end{eqnarray}
where to obtain the previous equations we have used the Bianchi identity, $\nabla_\mu G^{\mu\nu}=0$ and the second Bianchi identity $\nabla_\lambda R_{\alpha\beta\mu\nu}+\nabla_\mu R_{\alpha\beta\nu\lambda}+\nabla_\nu R_{\alpha\beta\lambda\mu}=0$.
Using the previous expressions and Eqs. (\ref{LReq}) and (\ref{sheareq}), one can see that Eq. (\ref{fieldvarphieq}) does not contain derivatives higher than two in both the metric and the scalar field $\varphi$ as expected. This is because the mimetic theory does not change the number of derivatives of the original theory \cite{Arroja:2015wpa} and in the previous case the original theory was Horndeski's theory which is well-known to contain no derivatives higher than two in the equations of motion. Note also that the equations of motion for $\lambda$ and $\varphi$, Eqs. (\ref{fieldlambdaeq}) and (\ref{fieldvarphieq}) respectively, are first-order ordinary differential equations. Following \cite{Lim:2010yk} one can argue that the Cauchy problem has a unique solution locally that depends only on the two initial conditions for $\varphi$ and $\lambda$. Furthermore from Eqs. (\ref{fieldlambdaeq}) and (\ref{fieldvarphieq}) one can see that the solutions evolve along time-like geodesics and neighbouring space points do not ``communicate" with each other, this implies that the sound speed is identically zero (for any cosmological background but with a non-dynamical metric) as we wanted to show. The Cauchy problem may become ill-defined for some time in the future for initial condition that give origin to caustics. This would be a problem for this model which is beyond the scope of this paper. Caustics are known to appear in other theories with non-canonical scalar fields, see for example \cite{Mukohyama:2009tp,Blas:2009yd}.

\section{Conclusions\label{sec:CON}}

In this paper we have studied linear scalar perturbations around a flat FLRW background in mimetic Horndeski gravity. This work is an important first step in the study of the evolution of cosmological perturbations in this very general class of mimetic models.
We have found that, in the absence of matter, the first-order equations of motion take a simple form given by Eqs. (\ref{seteom1})-(\ref{seteom3}) for the Bardeen potentials or equivalently Eqs. (\ref{seteomccp1})-(\ref{seteomccp3}) using the comoving curvature perturbation.
Just like in Horndeski's theory, in mimetic Horndeski gravity there is an effective anisotropic stress even in the absence of matter.

The (generalised) Newtonian potential was shown to satisfy a second-order ordinary differential equation with no spatial derivatives which implies that the sound speed for scalar perturbations is exactly zero for a flat FLRW background.
We have explicitly solved this equation for mimetic Horndeski models which include the so-called cubic Galileon term. This case includes the cosmological models proposed in our previous work \cite{Arroja:2015wpa}, where we showed that simple mimetic models can essentially mimic any desired expansion history. For this particular case, the form of the solutions that we found is the same as in the so-called $\lambda\varphi$ fluids \cite{Lim:2010yk} which are a generalisation of the mimetic dark matter scenario \cite{Chamseddine:2013kea,Chamseddine:2014vna}. We have shown that in these models, if the background expansion history is exactly equal to the $\Lambda$CDM expansion history then also the perturbations will evolve in exactly the same way.

The equation of motion for the comoving curvature perturbation is a first order ordinary differential equation which can be easily solved to find that the comoving curvature perturbation in general mimetic Horndeski gravity is exactly constant on all scales.

These results show that in these models there are no wave-like propagating scalar degrees of freedom.

We have shown that the conclusion that the sound speed of scalar perturbations is exactly zero around a flat FLRW background also applies to a mimetic theory beyond mimetic Horndeski, i.e. mimetic $G^3$ theories. (Mimetic) $G^3$ theories are interesting because they contain only one extra scalar degree of freedom in addition to the usual two polarizations of the graviton. On the other hand it is well-known that if one starts with a theory with higher-derivative equations of motion and which contains additional scalar degrees of freedom then the mimetic theory may have a non-zero sound speed \cite{Chamseddine:2014vna}. These results indicate that there might be a relation between the value of the scalar sound speed and the number of degrees of freedom of the original theory. We leave the detailed investigation of this relation for future work.

Finally, for a non-dynamical metric, we have shown that the scalar sound speed is exactly zero for all cosmological backgrounds.

For future work, we leave the study of the Hamiltonian formulation of mimetic Horndeski gravity and the issue of whether caustics will develop or not and if so how one can interpret them.

\section{Acknowledgments}

We would like to thank Dario Bettoni, Ryo Namba and Luigi Pilo for useful discussions and comments on an early version of this manuscript. Some computations in this paper were performed using Mathematica\footnote{https://www.wolfram.com/mathematica/}, the tensor computed algebra package xAct\footnote{http://www.xact.es/} and its contributed package xPand\footnote{http://www.xact.es/xPand/} \cite{Pitrou:2013hga}. FA is supported by the National Taiwan University (NTU) under Project No. 103R4000 and by the NTU Leung Center for Cosmology and Particle Astrophysics (LeCosPA) under Project No. FI121. PK acknowledges financial support from a Cariparo foundation grant.

\appendix
\section{The background equations of motion\label{BACK}}

In this appendix, we provide the expressions for the tensor $E_{\mu\nu}$ on a flat FLRW background. The same expressions can be used for the Horndeski and mimetic Horndeski models.
The non-zero components read
\begin{eqnarray}
E_{00}^{(0)}&=&
-a^2 K
-6 G_4 \mathcal{H}^2-6 G_{4,\varphi } \mathcal{H} \varphi_0'
+\left(\varphi _0'\right)^2 \left(\frac{12 G_{4,X} \mathcal{H}^2-9 G_{5,\varphi } \mathcal{H}^2}{a^2}-G_{3,\varphi }+K_{,X}\right)
\nonumber\\
&&+\left(\varphi_0'\right)^3 \left(\frac{5 G_{5,X} \mathcal{H}^3}{a^4}+\frac{3 G_{3,X} \mathcal{H}-6 G_{4,X\varphi} \mathcal{H}}{a^2}\right)
+\left(\varphi _0'\right)^4\frac{ \left(6 G_{4,XX} \mathcal{H}^2-3 G_{5,X\varphi }\mathcal{H}^2\right)}{a^4}
+\left(\varphi _0'\right)^5\frac{G_{5,XX} \mathcal{H}^3 }{a^6},
\\
E_{ij}^{(0)}&=&\delta _{ij } \bigg[
a^2 K 
+2 G_4 \mathcal{H}^2+4 G_4 \mathcal{H}'
+\varphi _0' \left(\frac{4 G_{5,\varphi } \mathcal{H} \varphi _0''-4 G_{4,X}\mathcal{H} \varphi _0''}{a^2}+2 G_{4,\varphi } \mathcal{H}\right)
\nonumber\\
&&+\left(\varphi _0'\right)^2 \bigg(-G_{3,\varphi }+2 G_{4,\varphi\varphi }-\frac{3 G_{5,X} \mathcal{H}^2 \varphi _0''}{a^4}
+\frac{-G_{3,X} \varphi_0''+2 G_{4,X}(\mathcal{H}^2-2\mathcal{H}')+2 G_{4,X\varphi } \varphi _0''+G_{5,\varphi }(2\mathcal{H}'-3\mathcal{H}^2)}{a^2}\bigg)
\nonumber\\
&&+\left(\varphi _0'\right)^3 \left(\frac{-4 G_{4,XX} \mathcal{H} \varphi _0''+2 G_{5,X\varphi } \mathcal{H} \varphi _0''+3 G_{5,X} \mathcal{H}^3-2 G_{5,X} \mathcal{H} \mathcal{H}'}{a^4}+\frac{G_{3,X} \mathcal{H}-6 G_{4,X\varphi} \mathcal{H}+2 G_{5,\varphi \varphi } \mathcal{H}}{a^2}\right)
\nonumber\\
&&+\left(\varphi _0'\right)^4 \bigg(-\frac{G_{5,XX} \mathcal{H}^2 \varphi _0''}{a^6}+\frac{4 G_{4,XX} \mathcal{H}^2-3 G_{5,X\varphi }\mathcal{H}^2}{a^4}\bigg)
+\frac{G_{5,XX} \mathcal{H}^3 \left(\varphi _0'\right)^5}{a^6} 
+2 G_{4,\varphi } \varphi _0''
   \bigg].\label{Eijzeroth}
\end{eqnarray}
The zeroth-order trace $E^{(0)}$ can be easily computed from the previous equations by using $E^{(0)}=-a^{-2}E_{00}^{(0)}+a^{-2}\delta^{ij}E_{ij}^{(0)}$.

\section{The explicit expressions of the $f_i$ functions\label{FI}}

In this appendix, we give the explicit expressions for the functions $f_i$, $i=1,...,21$ defined in the main text. These expressions can be used for both the Horndeski and mimetic Horndeski models because no equations of motion were used.

They read
\begin{eqnarray}
f_1&=&12 G_4 \mathcal{H}+6 G_{4,\varphi } \varphi _0'
 +\frac{ (18 G_{5,\varphi } \mathcal{H}-24 G_{4,X} \mathcal{H})}{a^2}\left(\varphi _0'\right)^2
 +\left(\varphi _0'\right)^3 \left(\frac{6 G_{4,X\varphi}-3 G_{3,X}}{a^2}-\frac{15 G_{5,X} \mathcal{H}^2}{a^4}\right)
 \nonumber\\
 &&+\frac{ (6G_{5,X\varphi} \mathcal{H}-12 G_{4,XX} \mathcal{H})}{a^4}\left(\varphi _0'\right)^4
 -\frac{3 G_{5,XX} \mathcal{H}^2 }{a^6}\left(\varphi _0'\right)^5,
\\
f_2&=&-6 G_{4,\varphi } \mathcal{H}
   +\varphi _0' \left(\frac{18 G_{4,X}\mathcal{H}^2-18 G_{5,\varphi } \mathcal{H}^2}{a^2}-2 G_{3,\varphi }+K_{,X}\right)
   +\left(\varphi _0'\right)^2 \left(\frac{15 G_{5,X} \mathcal{H}^3}{a^4}+\frac{9 G_{3,X}\mathcal{H}-24 G_{4,X\varphi} \mathcal{H}}{a^2}\right)
   \nonumber\\
   &&+\left(\varphi _0'\right)^3 \left(\frac{36G_{4,XX} \mathcal{H}^2-21 G_{5,X\varphi} \mathcal{H}^2}{a^4}+\frac{K_{,XX}-G_{3,X\varphi}}{a^2}\right)
    +\left(\varphi _0'\right)^4\left(\frac{10 G_{5,XX} \mathcal{H}^3}{a^6}+\frac{3 G_{3,XX} \mathcal{H}-6 G_{4,XX\varphi}\mathcal{H}}{a^4}\right)
    \nonumber\\
   &&+\frac{\left(6G_{4,XXX} \mathcal{H}^2-3 G_{5,XX\varphi} \mathcal{H}^2\right)}{a^6}\left(\varphi _0'\right)^5
   + \frac{G_{5,XXX} \mathcal{H}^3 }{a^8}\left(\varphi _0'\right)^6,
\\
f_3&=&-2 a^2 K
+\left(\varphi_0'\right)^2 \left(\frac{18 G_{5,\varphi } \mathcal{H}^2-18 G_{4,X} \mathcal{H}^2}{a^2}+K_{,X}\right)
+\left(\varphi _0'\right)^3 \left(\frac{18G_{4,X\varphi}\mathcal{H}-6 G_{3,X} \mathcal{H}}{a^2}-\frac{20 G_{5,X} \mathcal{H}^3}{a^4}\right)
\nonumber\\
&&+\left(\varphi _0'\right)^4\left(\frac{21 G_{5,X\varphi} \mathcal{H}^2-36 G_{4,XX} \mathcal{H}^2}{a^4}+\frac{G_{3,X\varphi}-K_{,XX}}{a^2}\right)
+\left(\varphi_0'\right)^5 \left(\frac{6 G_{4,XX\varphi} \mathcal{H}-3 G_{3XX} \mathcal{H}}{a^4}-\frac{11G_{5,XX} \mathcal{H}^3}{a^6}\right)
\nonumber\\
&&+\left(\varphi _0'\right)^6\frac{ \left(3G_{5,XX\varphi} \mathcal{H}^2-6 G_{4,XXX} \mathcal{H}^2\right)}{a^6}
-\frac{G_{5,XXX} \mathcal{H}^3 \left(\varphi _0'\right)^7}{a^8},
\\
f_4&=&-a^2 K_{,\varphi }-6 G_{4,\varphi } \mathcal{H}^2-6 G_{4,\varphi \varphi }\mathcal{H} \varphi _0'
+\left(\varphi_0'\right)^2 \left(\frac{12 G_{4,X\varphi} \mathcal{H}^2-9 G_{5,\varphi \varphi }\mathcal{H}^2}{a^2}-G_{3,\varphi \varphi }+K_{,X\varphi}\right)
\nonumber\\
&&+\left(\varphi _0'\right)^3\left(\frac{5 G_{5,X\varphi} \mathcal{H}^3}{a^4}+\frac{3 G_{3,X\varphi} \mathcal{H}-6G_{4,X\varphi\varphi} \mathcal{H}}{a^2}\right)
+\frac{ \left(6G_{4,XX\varphi} \mathcal{H}^2-3 G_{5,X\varphi\varphi} \mathcal{H}^2\right)}{a^4}\left(\varphi _0'\right)^4
+\frac{G_{5,XX\varphi} \mathcal{H}^3 }{a^6}\left(\varphi _0'\right)^5,
\nonumber\\
\\
f_5&=&-4 G_4+\frac{(4 G_{4,X}-2 G_{5,\varphi }) \left(\varphi_0'\right)^2}{a^2}+\frac{2 G_{5,X} \mathcal{H} \left(\varphi _0'\right)^3}{a^4},
\\
f_6&=&2 G_{4,\varphi }+\frac{\varphi _0' (4 G_{5,\varphi }\mathcal{H}-4 G_{4,X} \mathcal{H})}{a^2}+\left(\varphi _0'\right)^2 \left(\frac{2 G_{4,X\varphi}-G_{3,X}}{a^2}-\frac{3 G_{5,X} \mathcal{H}^2}{a^4}\right)+\frac{ (2G_{5,X\varphi} \mathcal{H}-4 G_{4,XX} \mathcal{H})}{a^4}\left(\varphi _0'\right)^3
\nonumber\\
&&-\frac{G_{5,XX} \mathcal{H}^2 \left(\varphi _0'\right)^4}{a^6},
\\
f_7&=&-2 G_4+\left(\varphi _0'\right)^2 \left(\frac{G_{5,X} \varphi_0''}{a^4}+\frac{G_{5,\varphi }}{a^2}\right)-\frac{G_{5,X} \mathcal{H} }{a^4}\left(\varphi _0'\right)^3,
\\
f_8&=&2 G_{4,\varphi }
+\left(\varphi _0'\right)^2 \left(\frac{(G_{5,X\varphi}-2 G_{4,XX}) \varphi_0''}{a^4}+\frac{2 G_{5,X} \mathcal{H}^2-G_{5,X} \mathcal{H}'}{a^4}+\frac{G_{5,\varphi \varphi }-2 G_{4,X\varphi}}{a^2}\right)
\nonumber\\
&&+\left(\varphi _0'\right)^3 \left(\frac{2G_{4,XX} \mathcal{H}-2 G_{5,X\varphi} \mathcal{H}}{a^4}-\frac{G_{5,XX} \mathcal{H} \varphi_0''}{a^6}\right)
+\frac{G_{5,XX} \mathcal{H}^2 }{a^6}\left(\varphi _0'\right)^4
-\frac{2 G_{5,X} \mathcal{H} }{a^4}\varphi _0' \varphi_0''
+\frac{(2 G_{5,\varphi }-2 G_{4,X}) }{a^2}\varphi _0'',\nonumber\\
\\
f_9&=&2 G_4+\frac{(G_{5,\varphi }-2 G_{4,X}) }{a^2}\left(\varphi_0'\right)^2-\frac{G_{5,X} \mathcal{H} }{a^4}\left(\varphi _0'\right)^3,
\\
f_{11}&=&2 G_{4,\varphi }
+\frac{ (4 G_{5,\varphi }\mathcal{H}-4 G_{4,X} \mathcal{H})}{a^2}\varphi _0'
+\left(\frac{2 G_{4,X\varphi}-G_{3,X}}{a^2}-\frac{3 G_{5,X} \mathcal{H}^2}{a^4}\right)\left(\varphi _0'\right)^2
+\frac{ (2G_{5,X\varphi} \mathcal{H}-4 G_{4,XX} \mathcal{H})}{a^4}\left(\varphi _0'\right)^3
\nonumber\\
&&-\frac{G_{5,XX} \mathcal{H}^2 \left(\varphi _0'\right)^4}{a^6},
\end{eqnarray}
\begin{eqnarray}
f_{12}&=&-8 G_4 \mathcal{H}+\varphi _0' \left(\frac{(4 G_{4,X}-4 G_{5,\varphi}) \varphi _0''}{a^2}-4 G_{4,\varphi }\right)
   +\left(\varphi_0'\right)^2 \left(\frac{6 G_{5,X} \mathcal{H} \varphi _0''}{a^4}+\frac{4 G_{4,X}\mathcal{H}}{a^2}\right)
\nonumber\\
&&+\left(\varphi _0'\right)^3 \left(\frac{(4 G_{4,XX}-2 G_{5,X\varphi}) \varphi_0''}{a^4}+\frac{2 G_{5,X} \mathcal{H}'-4 G_{5,X} \mathcal{H}^2}{a^4}+\frac{4 G_{4,X\varphi}-2 G_{5,\varphi\varphi }}{a^2}\right)
\nonumber\\
&&+\left(\varphi _0'\right)^4 \left(\frac{2 G_{5,XX} \mathcal{H}\varphi _0''}{a^6}+\frac{4 G_{5,X\varphi} \mathcal{H}-4 G_{4,XX} \mathcal{H}}{a^4}\right)
   -\frac{2 G_{5,XX} \mathcal{H}^2 }{a^6}\left(\varphi _0'\right)^5,
   \\
f_{13}&=&2 G_{4,\varphi } \mathcal{H}
\nonumber\\
&&+\varphi _0' \left(\frac{2 G_{4,X} (3\mathcal{H}^2-2 \mathcal{H}')-2G_{5,\varphi }(3 \mathcal{H}^2-2 \mathcal{H}')}{a^2}+\varphi _0'' \left(\frac{6 G_{4,X\varphi}-2G_{3,X}}{a^2}-\frac{6 G_{5,X} \mathcal{H}^2}{a^4}\right)-2 G_{3,\varphi }+4 G_{4,\varphi \varphi}+K_{,X}\right)
\nonumber\\
&&+\left(\varphi _0'\right)^2 \left(\frac{\varphi _0'' (10 G_{5,X\varphi} \mathcal{H}-16G_{4,XX} \mathcal{H})}{a^4}+\frac{9 G_{5,X} \mathcal{H}^3-6 G_{5,X} \mathcal{H} \mathcal{H}'}{a^4}+\frac{3G_{3,X} \mathcal{H}-16 G_{4,X\varphi} \mathcal{H}+6 G_{5,\varphi \varphi }\mathcal{H}}{a^2}\right)
\nonumber\\
&&+\left(\varphi _0'\right)^3 \bigg(\frac{18 G_{4,XX} \mathcal{H}^2-4 G_{4,XX}\mathcal{H}'-15 G_{5,X\varphi} \mathcal{H}^2+2 G_{5,X\varphi} \mathcal{H}'}{a^4}+\frac{2
   G_{4,X\varphi\varphi}-G_{3,X\varphi}}{a^2}
\nonumber\\
   &&+\varphi _0'' \left(\frac{2 G_{4,XX\varphi}-G_{3,XX}}{a^4}-\frac{7 G_{5,XX} \mathcal{H}^2}{a^6}\right)\bigg)
\nonumber\\
   &&+\left(\varphi _0'\right)^4 \left(\frac{\varphi _0'' (2 G_{5,XX\varphi} \mathcal{H}-4G_{4,XXX} \mathcal{H})}{a^6}+\frac{8 G_{5,XX} \mathcal{H}^3-2 G_{5,XX}\mathcal{H}\mathcal{H}'}{a^6}+\frac{G_{3,XX} \mathcal{H}-6 G_{4,XX\varphi} \mathcal{H}+2 G_{5,X\varphi\varphi} \mathcal{H}}{a^4}\right)
\nonumber\\
&&+\left(\varphi _0'\right)^5 \left(\frac{4 G_{4,XXX} \mathcal{H}^2-3G_{5,XX\varphi} \mathcal{H}^2}{a^6}-\frac{G_{5,XXX} \mathcal{H}^2 \varphi_0''}{a^8}\right)
   +\frac{G_{5,XXX} \mathcal{H}^3 \left(\varphi _0'\right)^6}{a^8}
+\frac{\varphi _0'' (4 G_{5,\varphi } \mathcal{H}-4G_{4,X} \mathcal{H})}{a^2},
\\
f_{14}&=&
-4 G_4 \mathcal{H}-2 G_{4,\varphi } \varphi _0'
+\left(\varphi _0'\right)^2\frac{ (8 G_{4,X}\mathcal{H}-6 G_{5,\varphi } \mathcal{H})}{a^2}
+\left(\varphi _0'\right)^3 \left(\frac{5 G_{5,X}\mathcal{H}^2}{a^4}+\frac{G_{3,X}-2 G_{4,X\varphi}}{a^2}\right)
\nonumber\\
&&+\left(\varphi _0'\right)^4\frac{ (4G_{4,XX} \mathcal{H}-2 G_{5,X\varphi} \mathcal{H})}{a^4}
+\frac{G_{5,XX} \mathcal{H}^2 \left(\varphi _0'\right)^5}{a^6},
\\
f_{16}&=&a^2 K_{,\varphi }
+2 G_{4,\varphi } \mathcal{H}^2+4 G_{4,\varphi }\mathcal{H}'
+\varphi _0' \left(\frac{\varphi _0'' (4G_{5,\varphi \varphi } \mathcal{H}-4 G_{4,X\varphi} \mathcal{H})}{a^2}+2 G_{4,\varphi \varphi }\mathcal{H}\right)
\nonumber\\
   &&+\left(\varphi _0'\right)^2 \left(\frac{2 G_{4,X\varphi} (\mathcal{H}^2-2 \mathcal{H}')+G_{5,\varphi \varphi }(2\mathcal{H}'-3 \mathcal{H}^2)}{a^2}+\varphi _0'' \left(\frac{2 G_{4,X\varphi\varphi}-G_{3,X\varphi}}{a^2}-\frac{3
   G_{5,X\varphi} \mathcal{H}^2}{a^4}\right)-G_{3,\varphi \varphi }+2 G_{4,\varphi \varphi \varphi
   }\right)
   \nonumber\\
   &&+\left(\varphi _0'\right)^3 \left(\frac{\varphi _0'' (2 G_{5,X\varphi\varphi} \mathcal{H}-4
   G_{4,XX\varphi} \mathcal{H})}{a^4}+\frac{3 G_{5,X\varphi} \mathcal{H}^3-2 G_{5,X\varphi}
   \mathcal{H} \mathcal{H}'}{a^4}+\frac{G_{3,X\varphi} \mathcal{H}-6 G_{4,X\varphi\varphi}
   \mathcal{H}+2 G_{5,\varphi \varphi \varphi } \mathcal{H}}{a^2}\right)
   \nonumber\\
&&+\left(\varphi _0'\right)^4 \left(\frac{4 G_{4,XX\varphi}
   \mathcal{H}^2-3 G_{5,X\varphi\varphi} \mathcal{H}^2}{a^4}-\frac{G_{5,XX\varphi} \mathcal{H}^2
   \varphi _0''}{a^6}\right)+\frac{G_{5,XX\varphi} \mathcal{H}^3 \left(\varphi _0'\right)^5}{a^6}
   +2 G_{4,\varphi \varphi } \varphi _0'',
\\
f_{17}&=&-4 G_4\mathcal{H}^2-8 G_4 \mathcal{H}'
 +\varphi _0' \left(\frac{\varphi _0'' (16 G_{4,X}
   \mathcal{H}-16 G_{5,\varphi } \mathcal{H})}{a^2}-4 G_{4,\varphi } \mathcal{H}\right)
\nonumber\\
   &&+\left(\varphi _0'\right)^2 \bigg(+2 G_{3,\varphi }-4
   G_{4,\varphi \varphi }-K_{,X}+\frac{-10 G_{4,X}
   \mathcal{H}^2+12 G_{4,X} \mathcal{H}'+12 G_{5,\varphi } \mathcal{H}^2-8 G_{5,\varphi } \mathcal{H}'}{a^2}
   \nonumber\\
   &&+\varphi _0''
   \left(\frac{18 G_{5,X} \mathcal{H}^2}{a^4}+\frac{4 G_{3,X}-10 G_{4,X\varphi}}{a^2}\right)\bigg)
   \nonumber\\
   &&+\left(\varphi _0'\right)^3 \left(\frac{\varphi _0'' (28 G_{4,XX}
   \mathcal{H}-16 G_{5,X\varphi} \mathcal{H})}{a^4}+\frac{12 G_{5,X} \mathcal{H} \mathcal{H}'-18 G_{5,X}
   \mathcal{H}^3}{a^4}+\frac{-4 G_{3,X} \mathcal{H}+22 G_{4,X\varphi} \mathcal{H}-8 G_{5,\varphi \varphi }
   \mathcal{H}}{a^2}\right)
   \nonumber\\
   &&+\left(\varphi _0'\right)^4 \bigg(\frac{-26 G_{4,XX} \mathcal{H}^2+4 G_{4,XX}
   \mathcal{H}'+21 G_{5,X\varphi} \mathcal{H}^2-2 G_{5,X\varphi}
   \mathcal{H}'}{a^4}+\frac{G_{3,X\varphi}-2 G_{4,X\varphi\varphi}}{a^2}\nonumber\\
   &&+\varphi _0'' \left(\frac{11
   G_{5,XX} \mathcal{H}^2}{a^6}+\frac{G_{3,XX}-2 G_{4,XX\varphi}}{a^4}\bigg)\right)
   \nonumber\\
   &&+\left(\varphi _0'\right)^5 \left(\frac{\varphi _0'' (4 G_{4,XXX}
   \mathcal{H}-2 G_{5,XX\varphi} \mathcal{H})}{a^6}+\frac{2 G_{5,XX} \mathcal{H} \mathcal{H}'-11
   G_{5,XX} \mathcal{H}^3}{a^6}+\frac{-G_{3,XX} \mathcal{H}+6 G_{4,XX\varphi} \mathcal{H}-2
   G_{5,X\varphi\varphi} \mathcal{H}}{a^4}\right)
   \nonumber \\
   &&+\left(\varphi _0'\right)^6
   \left(\frac{G_{5,XXX} \mathcal{H}^2 \varphi _0''}{a^8}+\frac{3 G_{5,XX\varphi} \mathcal{H}^2-4
   G_{4,XXX} \mathcal{H}^2}{a^6}\right)
   -\frac{G_{5,XXX} \mathcal{H}^3 }{a^8}\left(\varphi _0'\right)^7
   -4 G_{4,\varphi } \varphi _0'',
\\
f_{20}&=&
-2 G_{4,\varphi } \mathcal{H}
+\left(\frac{6 G_{4,X}\mathcal{H}^2-6 G_{5,\varphi } \mathcal{H}^2}{a^2}-2 G_{3,\varphi }+2 G_{4,\varphi \varphi }+K_{,X}\right)\varphi _0'
\nonumber\\
&&+ \left(\frac{3 G_{5,X} \mathcal{H}^3}{a^4}+\frac{3 G_{3,X} \mathcal{H}-10 G_{4,X\varphi}\mathcal{H}+2 G_{5,\varphi \varphi } \mathcal{H}}{a^2}\right)\left(\varphi_0'\right)^2
+\frac{ \left(6G_{4,XX} \mathcal{H}^2-4 G_{5,X\varphi} \mathcal{H}^2\right)}{a^4}\left(\varphi _0'\right)^3
+\frac{G_{5,XX} \mathcal{H}^3 }{a^6}\left(\varphi _0'\right)^4.\nonumber\\
\end{eqnarray}

The functions $f_i$ obey the following identities:
\begin{eqnarray}
&&f_{10}=f_{18},\qquad f_{11}=f_{19}, \qquad f_{21}=f_{14}, \qquad f_9=-\frac{f_{10}}{2}, \qquad\frac{f_{10}'-f_{12}}{f_{10}}=-2\mathcal{H},
\\
&&f_{11}(f_{10}'-f_{12})+f_{10}(f_{13}-f_{20}-f_{11}')=0, \qquad f_{14}-\mathcal{H}f_{10}+\varphi_0'f_{11}=0,
\\
&&f_{17}-\frac{f_{15}f_9}{f_7}-\frac{f_{14}}{f_{10}}(f_{12}-f_{10}')-f_{14}'+3E_{ij}^{(0)}+E_{00}^{(0)}+\alpha E_{ij}^{(0)}=0,
\\
&&\left(f_{16}-\frac{f_8f_{15}}{f_7}-\frac{f_{20}}{f_{10}}(f_{12}-f_{10}')-f_{20}'\right)\varphi_0'+E_{00}^{(0)'}+\mathcal{H}(3E_{ij}^{(0)}+E_{00}^{(0)})+\beta E_{ij}^{(0)}=0,
\\
&&\left(\frac{\mathcal{H}}{\varphi_0'}\right)'f_{10}+\left(\frac{\varphi_0''}{(\varphi_0')^2}-\frac{\mathcal{H}}{\varphi_0'}\right)f_{14}-f_{20}-a^2\frac{E^{(0)}}{\varphi_0'}+4\frac{E_{ij}^{(0)}}{\varphi_0'}=0,\qquad 2E_{ij}^{(0)}=-f_{15},
\end{eqnarray}
where
\begin{equation}
\alpha=-2-2\frac{f_9}{f_7}=-\frac{2(\varphi_0')^2G4_{,X}}{a^2G_4}
+
\frac{2 (\varphi_0')^2 \left(\left(a^2 {G_5}_{,\varphi}-\mathcal{H}
{G_5}_{,X} \varphi_0'\right) \left(2 a^2 {G_4}-{G_4}_{,X}
(\varphi_0')^2\right)+{G_5}_{,X} \left(a^2 {G_4}-{G_4}_{,X}
(\varphi_0')^2\right) \varphi_0''\right)}{a^2 {G_4} \left(2 a^4
{G_4}-a^2 {G_5}_{,\varphi} (\varphi_0')^2+{G_5}_{,X} (\varphi_0')^2
\left(\mathcal{H} \varphi_0'-\varphi_0''\right)\right)},
\end{equation}
and
\begin{eqnarray}
\beta&=&-2\varphi_0'\frac{f_8}{f_7}=\frac{2 \varphi_0' \left(a^4 {G_4}_{,\varphi}+{G_4}_{,XX}
(\varphi_0')^2 \left(\mathcal{H} \varphi_0'-\varphi_0''\right)-a^2
\left({G_4}_{,X\varphi} (\varphi_0')^2+{G_4}_{,X} \varphi_0''\right)\right)}{a^4 {G_4}}
\nonumber\\
&&+
\frac{2 \varphi_0'}{a^4 {G_4} \left(2 a^4
{G_4}-a^2 {G_5}_{,\varphi} (\varphi_0')^2+{G_5}_{,X} (\varphi_0')^2
\left(\mathcal{H} \varphi_0'-\varphi_0''\right)\right)}
\nonumber\\
&&\times
\Bigg[-{G_4}_{,XX} {G_5}_{,X} (\varphi_0')^4
\left(-\mathcal{H} \varphi_0'+\varphi_0''\right)^2
+a^6 \Big(\left({G_4}_{,\varphi}
{G_5}_{,\varphi}+{G_4} {G_5}_{\varphi\varphi}\right) (\varphi_0')^2+2
{G_4} {G_5}_{,\varphi} \varphi_0''\Big)
\nonumber\\&&
\qquad
+a^2
(\varphi_0')^2 \left(\mathcal{H} \varphi_0'-\varphi_0''\right)
\Big({G_4} \mathcal{H} {G_5}_{,XX}
\varphi_0'+\left({G_4}_{,X\varphi} {G_5}_{,X}+{G_4}_{,XX}
{G_5}_{,\varphi}\right) (\varphi_0')^2+{G_4}_{,X} {G_5}_{,X}
\varphi_0''\Big)
\nonumber\\&&
\qquad
+a^4 \varphi_0'
\Bigg(\varphi_0' \Big({G_4} {G_5}_{,X} \left(2
\mathcal{H}^2-\mathcal{H}'\right)-\mathcal{H} \left({G_4}_{,\varphi}
{G_5}_{,X}+2 {G_4} {G_5}_{,X\varphi}\right)
\varphi_0'-{G_4}_{,X\varphi} {G_5}_{,\varphi} (\varphi_0')^2\Big)
\nonumber\\&&
\qquad\qquad\qquad
+\Big(-2 {G_4} \mathcal{H} {G_5}_{,X}+\left({G_4}_{,\varphi}
{G_5}_{,X}+{G_4} {G_5}_{,X\varphi}-{G_4}_{,X} {G_5}_{,\varphi}\right)
\varphi_0'\Big) \varphi_0''\Bigg)\Bigg],
\end{eqnarray}
where $E_{ij}^{(0)}$ in the previous equations denotes the coefficient of $\delta_{ij}$ in $E_{ij}^{(0)}$ (i.e. the expression inside square brackets in Eq. (\ref{Eijzeroth})). The explicit expressions for the $f_i$ functions with $i=10,15,18,19,21$ can be found easily from the previous identities and the provided expressions for the other $f_i$ functions.

\section{Linear equations of motion in mimetic Horndeski gravity coupled to matter\label{MATTEREOM}}

In this appendix we briefly summarize well-known expressions for the linear scalar perturbations of the energy-momentum tensor in the Poisson gauge, see for example the review \cite{Malik:2008im}, and then we present the linear equations of motion in mimetic Horndeski gravity coupled to fluid matter. We assume a general energy-momentum tensor of matter than may contain anisotropic stress. We also assume that there is no direct coupling between this matter fluid and the mimetic scalar field $\varphi$.

The energy-momentum tensor of the fluid that we consider has the form
\begin{equation}
T_{\mu\nu}=(\rho+P)u_\mu u_\nu+Pg_{\mu\nu}+\pi_{\mu\nu},
\end{equation}
where $\rho$ is the energy density, $P$ the pressure and $\pi_{\mu\nu}$ is the anisotropic stress tensor. $\pi_{\mu\nu}$ vanishes for a perfect fluid or a minimally coupled scalar field, however it is non-zero for a non-minimally coupled scalar field and free-streaming neutrinos (or radiation). $u^\mu$ is the 4-velocity and note that the 4-velocity in this appendix is not related with the 4-velocity introduced in Sec. \ref{sec:SS}. $\pi_{\mu\nu}$ obeys $\pi_{\mu\nu}u^\mu=0$ and $\pi_{\mu}^\mu=0$.
We assume that the anisotropic stress is a first-order quantity and that the 4-velocity is defined so that $\pi_{00}=\pi_{0i}=0$ \cite{LandauLifshitz6,Weinberg:2008zzc} (this is the so-called energy frame), while the spatial part of $\pi_{\mu\nu}$ can be decomposed as
\begin{equation}
\pi_{ij}=a^2\left(\partial_i\partial_j\Pi-\frac{1}{3}\delta_{ij}\partial^2\Pi+\frac{1}{2}(\partial_i\Pi_j+\partial_j\Pi_i)+\Pi_{ij}\right),
\end{equation}
where the vector $\Pi_i$ obeys $\partial^i\Pi_i=0$ and the tensor $\Pi_{ij}$ obeys $\Pi_i^i=\partial^i\Pi_{ij}=0$ (where the indices are raised with $\delta^{ij}$). From now on we will neglect the vector and tensor parts of the anisotropic stress tensor.
The 4-velocity obeys the constraint $u_\mu u^\mu=-1$ and can be expanded as
\begin{equation}
u^0=a^{-1}(1-\Phi),\quad u^i=a^{-1}v^i,
\end{equation}
where the velocity $v^i$, a first-order quantity, can be decomposed in a scalar and intrinsic vector parts as $v^i=\delta^{ij}\partial_j v+v^i_\mathrm{vec}$, where $\partial_iv^i_\mathrm{vec}=0$. From now on we will also neglect $v^i_\mathrm{vec}$.
The zeroth-order components of the energy-momentum tensor are
\begin{equation}
T^{(0)}_{00}=a^2\rho_0,\quad T^{(0)}_{0i}=0,\quad T^{(0)}_{ij}=a^2P_0\delta_{ij},
\end{equation}
where $\rho_0$ and $P_0$ denote the zeroth-order energy density and pressure respectively. The trace is $T^{(0)}=-\rho_0+3P_0$.
At first order we have
\begin{eqnarray}
T^{(1)}_{00}=a^2\left(\delta\rho+2\rho_0\Phi\right),\quad T^{(1)}_{0i}=-a^2\left(\rho_0+P_0\right)\partial_iv,\quad T^{(1)}_{ij}=a^2\left((\delta P-2P_0\Psi)\delta_{ij}+\partial_i\partial_j\Pi-\frac{1}{3}\delta_{ij}\partial^2\Pi\right),
\end{eqnarray}
where $\delta\rho$ and $\delta P$ denote the energy density and pressure perturbations respectively. The trace is $T^{(1)}=-\delta\rho+3\delta P$.
The conservation of the energy-momentum tensor, $\nabla^\mu T_{\mu\nu}=0$ implies at zeroth order
\begin{equation}
\rho_0'+3\mathcal{H}(\rho_0+P_0)=0,
\end{equation}
and at first order
\begin{eqnarray}
&&
\delta\rho'+3\mathcal{H}(\delta\rho+\delta P)-3(\rho_0+P_0)\Psi'+(\rho_0+P_0)\partial^2v=0,
\\
&&
\left((\rho_0+P_0)v\right)'+\delta P+\frac{2}{3}\partial^2\Pi+4\mathcal{H}(\rho_0+P_0)v+(\rho_0+P_0)\Phi=0.
\end{eqnarray}
The previous results are all well-known in the literature and now we will present the equations of motion of mimetic Horndeski gravity coupled with this fluid.

The equations of motion of the mimetic Horndeski model including matter are Eqs. (\ref{mMF}) where the $E_{\mu\nu}$ tensor is computed from the Horndeski Lagrangian. They read
\begin{eqnarray}
&&
b(\varphi)g^{\mu\nu}\partial_\mu\varphi\partial_\nu\varphi-1=0,\\
&&
E^{\mu\nu}+T^{\mu\nu}=(E+T)b(\varphi)\partial^\mu\varphi\partial^\nu\varphi,\\
&&
\nabla_\mu T^{\mu\nu}=0,
\end{eqnarray}
where we dropped the field equation because it is redundant and replaced the equation $\Omega_m=0$ with the equivalent equation $\nabla_\mu T^{\mu\nu}=0$. As shown in section \ref{sec:MOD}, the ``time-time" component of the generalized Einstein equations is also redundant.
In the background the previous equations of motion reduce to
\begin{equation}
-a^{-2}b(\varphi_0)(\varphi_0')^2=1,\qquad E_{ij}^{(0)}=-a^2 P_0\delta_{ij}, \qquad \rho_0'+3\mathcal{H}(\rho_0+P_0)=0.
\end{equation}
At first order they are
\begin{eqnarray}
&&2b_0\delta\varphi'+\varphi_0'b_{,\varphi}\delta\varphi-2b_0\varphi_0'\Phi=0,\label{eq1}
\\
&&f_7\Psi+f_8\delta\varphi+f_9\Phi+a^2\Pi=0,\label{eq2}
\\
&&f_{10}\Psi''+f_{11}\delta\varphi''+f_{12}\Psi'+f_{13}\delta\varphi'+f_{14}\Phi'+f_{15}\Psi+f_{16}\delta\varphi+f_{17}\Phi+\frac{2}{3}a^2\partial^2\Pi+a^2\left(\delta P-2P_0\Psi\right)=0,\label{eq3}
\\
&&f_{10}\Psi'+f_{11}\delta\varphi'+\left(f_{20}+\frac{a^2(E^{(0)}+T^{(0)})}{\varphi_0'}\right)\delta\varphi+f_{14}\Phi-a^2\left(\rho_0+P_0\right)v=0,\label{eq4}
\\
&&\delta\rho'+3\mathcal{H}(\delta\rho+\delta P)-3(\rho_0+P_0)\Psi'+(\rho_0+P_0)\partial^2v=0,\label{eq5}
\\
&&
\left((\rho_0+P_0)v\right)'+\delta P+\frac{2}{3}\partial^2\Pi+4\mathcal{H}(\rho_0+P_0)v+(\rho_0+P_0)\Phi=0.\label{eq6}
\end{eqnarray}
Similarly to the case discussed in the main text, one can show that the third equation of the previous set can be derived from the other equations (one does not need to use the fifth equation to show that) and using the background equations of motion.
In conclusion the set of independent equations of motion in mimetic gravity with matter is given by Eqs. (\ref{eq1}), (\ref{eq2}), (\ref{eq4}), (\ref{eq5}) and (\ref{eq6}).
Using the variable $\zeta$ defined in subsection \ref{subsec:PERTMIMETICHORNDESKI} one can write Eq. (\ref{eq4}) in the previous set as
\begin{equation}
\zeta'=a^2\frac{\rho_0+P_0}{f_{10}}\left(\frac{\delta\varphi}{\varphi_0'}-v\right).
\end{equation}

\section{The sound speed in the mimetic $G^3$ theory\label{SSG3}}

Horndeski's theory is the most general 4D covariant scalar-tensor theory that can be derived from an action and contains only second order equations of motion. However it is known \cite{Gleyzes:2014dya,Gleyzes:2014qga,Gao:2014soa} that there are theories that include Horndeski's theory, can be derived from an action and are more general than Horndeski's theory. In some cases, these theories have been shown to propagate exactly the same number of degrees of freedom as Horndeski's theory and therefore are free from higher-derivative ghosts despite having covariant higher-order equations of motion. The theories presented in \cite{Gleyzes:2014dya,Gleyzes:2014qga} are also known as $G^3$ theories.

In this appendix we show that even in a mimetic $G^3$ theory (without matter) at first order around a flat FLRW background, the speed of sound of scalar perturbations is still exactly zero.
The beyond Horndeski theory of \cite{Gleyzes:2014dya,Gleyzes:2014qga} is defined by adding to the Horndeski action the following action
\begin{eqnarray}
S_{G^3}\!\!&=&\!\!\int\! d^4x\sqrt{-g}\Bigg[
A_1(X,\varphi)\Bigg(\!\!-2X\left((\Box\varphi)^2-\nabla_\mu\nabla_\nu\varphi\nabla^\mu\nabla^\nu\varphi\right)-2\left(\nabla^\mu\varphi\nabla^\nu\varphi\nabla_\mu\nabla_\nu\varphi\Box\varphi-\nabla^\mu\varphi\nabla_\mu\nabla_\nu\varphi\nabla_\lambda\varphi\nabla^\lambda\nabla^\nu\varphi\right)\!\!\Bigg)
\nonumber\\
&&\qquad\qquad\quad
+A_2(X,\varphi)\Bigg(\!\!-2X\left((\Box\varphi)^3-3\Box\varphi\nabla_\mu\nabla_\nu\varphi\nabla^\mu\nabla^\nu\varphi+2\nabla_\mu\nabla_\nu\varphi\nabla^\nu\nabla^\rho\varphi\nabla^\mu\nabla_\rho\varphi\right)
\nonumber\\
&&\qquad\qquad\qquad\qquad\qquad\quad
-3\big((\Box\varphi)^2\nabla_\mu\varphi\nabla^\mu\nabla^\nu\varphi\nabla_\nu\varphi-2 \Box\varphi\nabla_\mu\varphi\nabla^\mu\nabla^\nu\varphi\nabla_\nu\nabla_\rho\varphi\nabla^\rho\varphi
\nonumber\\
&&\qquad\qquad\qquad\qquad\qquad\quad
-\nabla_\mu\nabla_\nu\varphi\nabla^\mu\nabla^\nu\varphi\nabla_\rho\varphi\nabla^\rho\nabla^\lambda\varphi\nabla_\lambda\varphi+2\nabla_\mu\varphi\nabla^\mu\nabla^\nu\varphi\nabla_\nu\nabla_\rho\varphi\nabla^\rho\nabla^\lambda\varphi\nabla_\lambda\varphi\big)\!\!\Bigg)
\Bigg],
\end{eqnarray}
where the functions $A_1$ and $A_2$ are free functions of their two arguments and together with the four free functions present in an Horndeski theory they define the $G^3$ theory.

The first-order components of the new $E_{\mu\nu}$ tensor coming from the previous action are of the form
\begin{eqnarray}
\tilde E_{00}^{(1)}&=&g_1\Psi'+g_2\delta\varphi'+g_3\delta\varphi+g_4\Phi+g_5\partial^2\delta\varphi,
\\
\tilde E_{ij}^{(1)}&=&\partial_i\partial_j\left(g_{6}\delta\varphi'+g_{7}\delta\varphi\right)+\delta_{ij}\Big(-g_{6}\partial^2\delta\varphi'-g_{7}\partial^2\delta\varphi
\nonumber\\
&&\qquad\qquad
+g_{8}\Psi''+g_{9}\delta\varphi''+g_{10}\Psi'+g_{11}\delta\varphi'+g_{12}\Phi'+g_{13}\Psi+g_{14}\delta\varphi+g_{15}\Phi\Big),
\\
\tilde E_{0i}^{(1)}&=&\partial_i\left(g_{16}\Psi'+g_{17}\delta\varphi'+g_{18}\delta\varphi+g_{19}\Phi\right),
\end{eqnarray}
where the $g_i$ with $i=1,\dots,19$ are functions of $A_1$, $A_2$ and their derivatives. We do not write the explicit expressions for these functions because they are rather long and they are not important for our discussion regarding the value of the sound speed.

The first-order equations of motion for the mimetic $G^3$ model are determined only by $\tilde E_{ij}^{(1)}$ and $\tilde E_{0i}^{(1)}$ and their counterparts for the remaining Horndeski terms. In the absence of matter, the $i-j$ equation of motion is simply $E_{ij}^{(1)}+\tilde E_{ij}^{(1)}=0$. This implies two equations as
\begin{eqnarray}
&&f_7\Psi+(f_8+g_7)\delta\varphi+f_9\Phi+g_{6}\delta\varphi'=0,\label{k2EijeomG3}
\\
&&(f_{10}+g_{8})\Psi''+(f_{11}+g_{9})\delta\varphi''+(f_{12}+g_{10})\Psi'+(f_{13}+g_{11})\delta\varphi'+(f_{14}+g_{12})\Phi'+(f_{15}+g_{13})\Psi
\nonumber\\
&&
+(f_{16}+g_{14})\delta\varphi+(f_{17}+g_{15})\Phi=0.\label{EijeomG3}
\end{eqnarray}
In a general mimetic theory (and in particular also for mimetic $G^3$) the equation $E_{00}^{(1)}+\tilde E_{00}^{(1)}=0$ is replaced by the mimetic constraint, Eq. (\ref{seteom1}), which at first order does not contain any spatial derivatives. Because also Eqs. (\ref{k2EijeomG3}) and (\ref{EijeomG3}) do not contain spatial derivatives, the sound speed of the mimetic $G^3$ model has to be zero as in the mimetic Horndeski model.


\end{document}